\documentclass[
reprint,
prd,
twocolumn,
nofootinbib,
tightenlines %
floatfix,
 amsmath,
showpacs ,amssymb, aps,
superscriptaddress
]{revtex4-1}
\usepackage{amsmath,slashed}
\usepackage[left]{lineno}
\usepackage{units}
\usepackage{comment}
\usepackage{graphicx}
\usepackage{amsmath,amssymb}
\usepackage{support-caption}
\usepackage{subcaption}
\usepackage{bm}
\usepackage[shortlabels]{enumitem}
\usepackage{float}
\usepackage{soul}
\usepackage{natbib}
\usepackage{wasysym}
\usepackage{aas_macros}
\graphicspath{{images/}}
\RequirePackage{lineno}
\usepackage[usenames,dvipsnames,table]{xcolor}
\usepackage{appendix}
\usepackage{extdash}
\usepackage{colortbl}
\usepackage{tikz}

\usepackage{booktabs}

\usepackage[printwatermark]{xwatermark}

\definecolor{lightblue}{HTML}{1F77B4}
\definecolor{grey}{HTML}{808080}

\newcommand{\tc}{t_{\mathrm{col}}}
\newcommand{\tcu}{t_{\mathrm{col,+1\sigma}}}
\newcommand{\tcl}{t_{\mathrm{col,-1\sigma}}}
\newcommand{\tw}{t_{\mathrm{w}}}

\newcommand{\SPA}{School of Physics and Astronomy, Monash University, Vic 3800, Australia}
\newcommand{\OzGravMonash}{OzGrav: The ARC Centre of Excellence for Gravitational Wave Discovery, Clayton VIC 3800, Australia}
\newcommand{\AstroThreeD}{ARC Centre of Excellence for All Sky Astrophysics in 3 Dimensions (ASTRO 3D), Canberra, ACT 2611, Australia}

\newcommand{\msun}{\mathrm{M}_{\odot}}

\definecolor{Waveform1}{HTML}{1f77b4}
\definecolor{Waveform2}{HTML}{ff7f0e}
\definecolor{Waveform3}{HTML}{2ca02c}
\definecolor{Waveform4}{HTML}{d62728}
\definecolor{Waveform5}{HTML}{9467bd}
\definecolor{Waveform6}{HTML}{8c564b}
\definecolor{Waveform7}{HTML}{e377c2}
\definecolor{Waveform8}{HTML}{7f7f7f}
\definecolor{Waveform9}{HTML}{bcbd22}

\begin{document}

\title{Can we measure the collapse time of a post-merger remnant for a future GW170817-like event?}

\author{Paul J. Easter}
\email{paul.easter@monash.edu}
\affiliation{\SPA}
\affiliation{\OzGravMonash}
\author{Paul D. Lasky}
\email{paul.lasky@monash.edu}
\affiliation{\SPA}
\affiliation{\OzGravMonash}
\author{Andrew R. Casey}
\email{andrew.casey@monash.edu}
\affiliation{\SPA}
\affiliation{\AstroThreeD}

\pacs{
}
\begin{abstract}

    Measuring the collapse time of a binary neutron star merger remnant can inform the physics of extreme matter and improve modelling of short gamma-ray bursts and associated kilonova.
    The lifetime of the post-merger remnant directly impacts the mechanisms available for the jet launch of short gamma-ray bursts.
    We develop and test a method to measure the collapse time of post-merger remnants.
    We show that for a GW170817-like event at $\sim\!40$\,Mpc, a network of Einstein Telescope with Cosmic Explorer is required to detect collapse times of $\sim\!10$\,ms.
    For a two-detector network at A+ design sensitivity, post-merger remnants with collapse times of $\sim\!10\,\mathrm{ms}$ must be $\lesssim 10$\,Mpc to be measureable.
    This increases to $\sim\!18\Hyphdash*26$\,Mpc if we include the proposed Neutron star Extreme Matter Observatory (NEMO), increasing the effective volume by a factor of $\sim\!30$.

\end{abstract}
\maketitle
    \vspace{-0.25cm}
    \section{Introduction}
    \label{sec:introduction}

    Measuring the lifetimes of binary neutron star post-merger remnants can help probe extreme matter at high temperature and densities and narrow down the physics of short gamma-ray bursts~\cite[e.g.,][]{Zhang2019,Ciolfi2020a}.
    These remnants may promptly collapse into a black hole or form a hot, differentially-rotating neutron star.
    If the total mass of a remnant is between 1.2 and 1.5 times the maximum non-rotating neutron star mass (the Tolman-Oppenheimer-Volkov mass), then the remnant is known as a hypermassive neutron star~\cite{Tolman1939,Oppenheimer1939,Breu2016,Weih2018,Baumgarte2000}, which is expected to collapse to form a black hole in a timescale from milliseconds to seconds \cite{Paschalidis2012}. 
    For smaller masses, the differentially-rotating remnant will evolve into rigidly-rotating neutron star after the differential rotation is quenched.
    The rigidly-rotating remnant will either collapse to a black hole, or form a stable neutron star, depending on the remnant mass.
    See Ref.~\cite{Sarin2020b} for a recent review on the evolution of neutron star merger remnants.

    Determining the collapse times of hypermassive remnants can help narrow down the nature of the central engine for short gamma-ray-bursts. 
    Multi-messenger observations of binary neutron star merger GW170817 suggest that the remnant may have either collapsed to a black hole \cite[e.g.,][]{Metzger2018,Gill2019,Murguia-Berthier2020}, or formed a long-lived remnant~\cite[e.g.][]{Yu2018}.
    Measuring the collapse time of a remnant may help determine the characteristic timescales associated with short gamma-ray-bursts, aiding the selection of the central engine (for a review see \cite{Zhang2019}).
    Furthermore, measuring the collapse time of the post-merger remnant may help constrain the quenching mechanism and physics of the differential rotation, which may reveal indicators towards the relative contribution of radiative (gravitational waves and neutrino) and dissipative (viscous, resistive and magnetic braking) processes within the remnant.\par

    The direct detection of gravitational waves from future neutron star merger remnants presents a great opportunity to constrain the collapse time.
    Numerical-relativity simulations of merger remnants show gravitational waves predominantly emitted from the fundamental f-mode oscillation of the remnant~\cite{Zhuge1994,Stergioulas2011}.
    Gravitational waves emitted from this mode occur at $\sim\!1.8\Hyphdash*4$\,kHz~\cite{Takami2015,Bernuzzi2015}.
    No post-merger remnant was detected for GW170817 by the LIGO and Virgo collaboration due to lack of sensitivity of the detectors at these frequencies. 
    However, increased sensitivity of gravitational-wave instruments and targeted high-frequency detectors  may enable future detections of post-merger remnants~\cite[e.g.,][]{Martynov2019,NEMO2020}. \par
    
    In this paper, we assess how well future networks of gravitational-wave detectors can measure the collapse time of a post-merger remnant.
    We extend a waveform model developed in Ref.~\cite{Easter2020} which was derived from Refs.~\cite{Bauswein2016,Bose2018}, to allow the measurement of the collapse time of the post-merger remnant.
    Using Bayesian inference, we inject numerical-relativity gravitational waveforms that are forced to collapse into different interferometer configurations to measure the maximum distance at which we can recover the collapse time.\par

    This paper is laid out as follows. In Section~\ref{sec:methodology} we outline the extended waveform model along with the method of gravitational-wave injections.
    The results are outlined in Section~\ref{sec:results} and their implications in Section~\ref{sec:discussion}.
    \vspace{-0.25cm}
    \section{methodology}\label{sec:methodology}
    We use numerical-relativity waveforms with two different equations of state that we inject into Gaussian noise realisations determined by the gravitational-wave interferometer configuration.
    We modify the numerical-relativity waveforms to collapse at a given collapse time.
    We then use an analytic model to perform detection and parameter estimation to determine distributions of the model parameters.
    This model, based on two models in Refs.~\cite{Bauswein2016,Bose2018}, is outlined in Ref.~\cite{Easter2020}.
    See also Refs.~\cite{Tsang2019,Breschi2019} for alternative models of the post-merger gravitational-wave strain.

\begin{figure*}[t]
        \centering
        \includegraphics[scale=0.35]{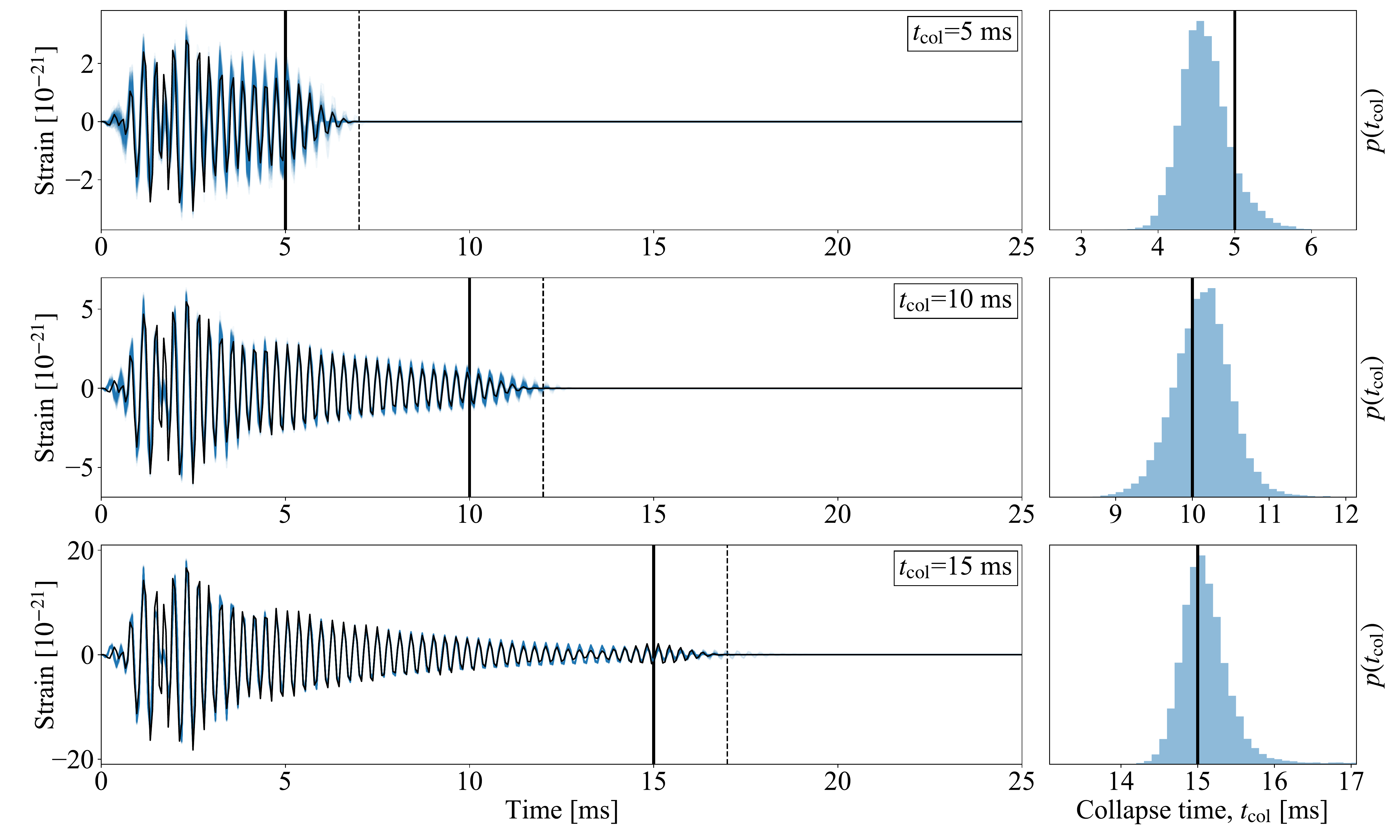}
        \caption{Time domain posterior waveforms for SLy equation of state with equal mass 1.35M${_\odot}$ progenitor (left panels) and measured collapse times (right panels) for injections with collapse times $\tc=5$\,ms~(top), $\tc=10$\,ms~(centre), and $\tc=15$\,ms~(bottom) injected into 2A+ detector network. Left panels show the numerical-relativity injections that have been forced to collapse (black curves) with the posterior waveforms (shaded blue curves). The right panels show the collapse-time posterior distributions for each numerical-relativity injection.
        The solid vertical lines on all panels shows  $\tc$ for the injected waveform. The dashed vertical lines shows  $\tc+\tw$ where the injected signal drops to zero due to the induced collapse. Here we show three loud injections from our injection study where $\tc$ are clearly recovered. The injected distances are 5.93\,Mpc, 3.04\,Mpc, and 1.00\,Mpc, for 5\,ms,  10\,ms, and  15\,ms respectively. The full posteriors for the $\tc=10\,$ms collapse time   injection are shown in the appendix (Fig~\ref{fig:FullCorner}).} 
        \label{fig:TimeResponse}
\end{figure*}     
\begin{figure*}[t]
    \centering
    \includegraphics[scale=0.29]{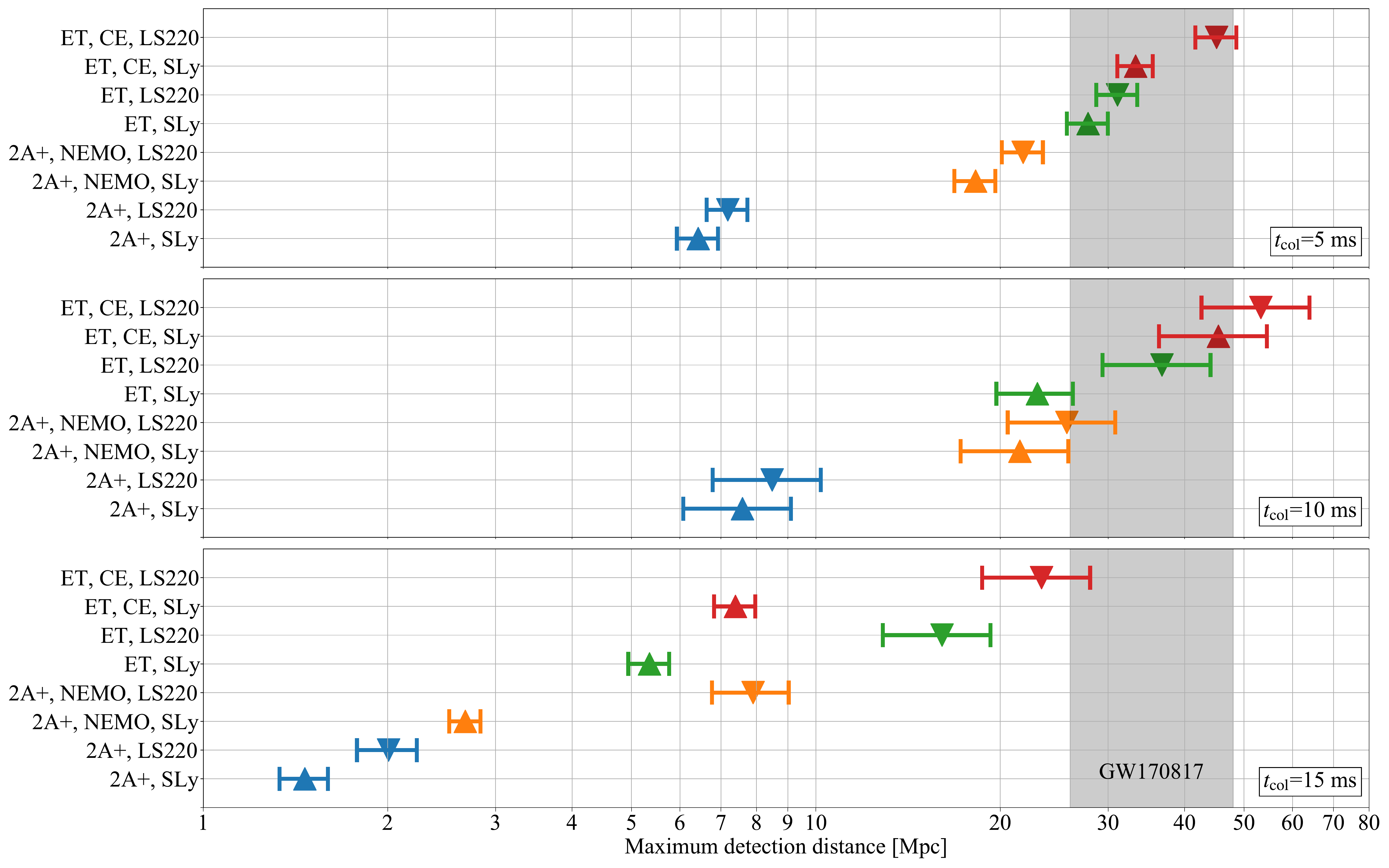}
    \caption{Maximum distances for which the collapse time can be measured for different interferometer networks and equations of state. The vertical axis shows different equations of state and interferometer configurations, and the maximum detection distance is shown on the horizontal axis. The numerical-relativity simulations are injected with equations of state SLy (upward pointing triangle) and LS220 (downward pointing triangle). The interferometer configurations are 2A+ (blue), 2A+ with the proposed Neutron star Extreme Matter Observatory (orange), Einstein Telescope (green), and Einstein Telescope with Cosmic Explorer~(red). The top panel shows collapse times of 5\,ms, the centre panel 10\,ms and the lower panel 15\,ms collapse times. The  luminosity distance for GW170817 (gravitational-wave only) is shown in shaded grey for comparison~\cite{GW170817multi}.} 
    \label{fig:CollapseTimeDistanceGrid}
\end{figure*}  
        The gravitational waves from the post-merger remnant are modelled as a third-order, exponentially-damped sinusoid with a linear frequency-drift term. The plus polarisation of the gravitational-wave strain, $h_{+}(\boldsymbol{\theta},t)$, is given by: 
\begin{align}
        h_{+}(\boldsymbol{\theta},t) & = \sum_{j} h_{j,+}(\boldsymbol{\theta},t),\nonumber\\
        h_{j,+}(\boldsymbol{\theta},t) & = 
        H w_j \exp\left[-\frac{t}{T_j}\right] \cos \left(2\pi f_j t\left[1+\alpha_j t\right]+\psi_j \right),\nonumber\\
        \boldsymbol{\theta} & = \{H,w_j,T_j,f_j,\alpha_j,\psi_j :  j \in [0,2] \},\label{eq:hplus}
\end{align}
    where $\boldsymbol{\theta}$ are the model parameters and $\sum_{j} w_j=1$. 
    Here, $H$ is an overall amplitude scaling factor, $w_j$ is the relative amplitude of the $j$th mode, $T_j$ is the corresponding exponential damping time constant, and $f_j$ is the corresponding frequency.
    The linear frequency-drift term is $\alpha_j$ and the initial phase  is $\psi_j$.
    The time, $t$, is defined, such that $\max\left[h_+^2(\boldsymbol{\theta},t)+h_\times^2(\boldsymbol{\theta},t)\right]$ occurs when $t=0$. 
    The cross polarisation for Eq.~\ref{eq:hplus} is found by applying a $\pi/2$ phase shift to $\psi_j$.
    More details on this model can be found in Ref.~\cite{Easter2020}.
    
    We extend this model by introducing a collapse in the gravitational-wave strain using the falling edge of a Tukey window.
    The falling edge starts at the collapse time, $\tc$, and by the time $t=\tc+\tw$ the gravitational-wave strain drops to zero (see Fig.~\ref{fig:TimeResponse}). 
    Here $\tw$ is the time taken for the gravitational-wave signal to completely decay.
    We have chosen $\tw=2$\,ms from examining numerical-relativity simulations with collapsing remnants (e.g., simulation labels: \texttt{BAM:0109:R01}, \texttt{BAM:0110:R01}, \texttt{BAM:0111:R02}~\cite{Dietrich2018}, \texttt{BAM:0044:R02}, \texttt{BAM:0045:R01}~\cite{Dietrich2017c}, \texttt{BAM:0124:R01}~\cite{Dietrich2017}).
    This model captures the reduction in amplitude of the post-merger gravitational-wave signal as the remnant collapses.
    The full gravitational-wave strain including the collapse of the remnant, $h_{\varphi,\mathrm{c}}(\boldsymbol{\tilde{\theta}},t)$, is given by:
\begin{align}
        & h_{\varphi,\mathrm{c}}(\boldsymbol{\tilde{\theta}},t)    =  h_{\varphi}(\boldsymbol{\theta},t)\mathcal{T}\left(t , \tc,\tw\right), \label{eq:hcol}\\ 
        \mathcal{T} & (t,\tc,\tw) =    \nonumber  \\
        &\begin{cases} 
            1 &\text{if}\ t \leqslant  \tc \\ 
            \frac{1}{2}\left(\cos\left(\frac{\pi (t-\tc)}{\tw}\right) +1 \right)  &\text{if}\ \tc < t \leqslant \tc+\tw\ \text{.} \label{eq:tukeyfallingedge} \\
            0  &\text{if}\  t > \tc+\tw\\ 
        \end{cases}
\end{align}
    Here, $\varphi\in\{+,\times\}$ are the plus and cross gravitational-wave polarisations, and $\boldsymbol{\tilde{\theta}}= \boldsymbol{\theta} \cup \{\tc,\tw\}$. \par
    
    To generate the collapsing gravitational-wave signal injection, $s_{\varphi,\mathrm{c}}(t)$, we apply the Tukey collapse window to the numerical-relativity simulation, $h_{\mathrm{NR},\varphi}(t)$, as follows: 
    \begin{align}
        s_{\varphi,\mathrm{c}}(t) & =  h_{\mathrm{NR},\varphi}(t)\mathcal{T}\left(t , \tc,\tw\right). \label{eq:scol}
    \end{align}
    To ensure that $s_{\varphi,\mathrm{c}}(t)$ has the required collapse time, $h_{\mathrm{NR},\varphi}(t)$ must emit post-merger gravitational waves for $t>\tc+\tw$.
    In this paper we use two numerical-relativity simulations of binary neutron star mergers with equal mass $1.35\,\mathrm{M}_\odot$ progenitors that emit gravitational waves for $\sim~25$\,ms and sample collapse times of 5, 10, and 15\,ms.
    The two simulations use the SLy equation of state~\cite{Douchin2001} (simulation label \texttt{THC:0036:R03} from Refs.~\cite{Dietrich2018,Radice2016}) and LS220 equation of state~\cite{Lattimer1991} (simulation label \texttt{THC:0019:R05} from Refs.~\cite{Dietrich2018,Radice2017}).
    The dimensionless tidal-deformabilities of the progenitor neutron stars for the two simulations are 390 and 684 respectively, which are consistent with tidal deformabilities inferred from GW170817~\cite{GW170817Properties}.
    The SLy equation of state is the softer of the two equations of state, with a lower tidal deformability, more compact remnant structure, and higher dominant oscillation frequency.\par
    
    We use four different detector networks in this injection study.
    Firstly, a two detector network at A+ design sensitivity (2A+) located at existing LIGO sites: Hanford, Washington; and Livingston, Louisiana~\cite{Abbott2020Prospects,PSD:Aplus}. 
    Secondly, we add the proposed Neutron star Extreme Matter Observatory (NEMO), located at Gingin, Western Australia to the first network~\cite{NEMO2020}.
    Thirdly, we use the Einstein Telescope (ET)~\cite{Hild2008,Punturo2010,Hild2011, PSD:ET}.
    And finally, the Einstein Telescope with an additional interferometer, located at Hanford, Washington, with a Cosmic Explorer (CE) power spectral density~\cite{Abbott2017b, Adhikari2019,PSD:CE}.
    We use fixed random seeds for Gaussian noise generation for each interferometer and inject numerical-relativity simulations at a sky position corresponding to a mean sky signal-to-noise ratio. 
    \par
      
    We use \textsc{Bilby}, a Bayesian inference package~\cite{Ashton2019}, to obtain posteriors, $p(\boldsymbol{\tilde{\theta}}~|~s_{\varphi,\mathrm{c}}(t))$ from a numerical relativity injection with enforced collapse starting at $t=\tc$ with width $\tw=2$\,ms.
    Posteriors are calculated using waveforms detailed in Eqs.~\ref{eq:hplus}\Hyphdash*\ref{eq:tukeyfallingedge}. 
    See Ref.~\cite{Easter2020} for further details of the gravitational-wave likelihood, and Appendix~\ref{appendix:a} for additional information on the priors.
    \par
    To deem the collapse time as successfully recovered, we demand the following to hold: 
\begin{align}
    \tcl - 2\,\mathrm{ms} \leqslant \tc & \leqslant  \tcu + 2\,\mathrm{ms},\\
    \tcu - \tcl & \leqslant  5\,\mathrm{ms},
\end{align}
    where $\tc$ is the injected value. 
    Here $\tcu$ and $\tcl$ are the upper and lower 68 percentile credible intervals on the posterior,  $p(\tc)$.
    Finally, the Bayes Factor for the ratio of evidence for signal against evidence for noise must be $\gg 1$.
    The minimum successful Bayes Factor in favour of a signal over noise in this paper is $\sim\!500$.
    These requirements ensure that successfully recovered $\tc$ are within a few milliseconds of the  injected values. 
    Although this method is somewhat arbitrary, it successfully identifies injections where the collapse time is recovered.
    Furthermore, as the results are pessimistic, with detections not expected until Cosmic Explorer and Einstein Telescope are online, changing this selection criterior will not substantially change these results. \par

    We perform numerical-relativity injections with the full waveform, Eq.~\ref{eq:scol}, at a grid of distances and apply the above rules to determine whether we successfully recover $\tc$.

\vspace{-0.25cm}    
\section{Results}\label{sec:results}
    We inject post-merger numerical-relativity waveforms modified to collapse at $t=\tc$ and sample posteriors for $\boldsymbol{\tilde{\theta}}$.
    We then calculate posterior waveforms, $h_{\varphi,c}(\boldsymbol{\tilde{\theta}},t)$, from Eqs.~\ref{eq:hcol}\Hyphdash*\ref{eq:tukeyfallingedge}.
    Fig.~\ref{fig:TimeResponse} shows example posterior waveforms for the plus polarisation (left panels) and collapse-time posterior distributions (right panels) for numerical-relativity injections using an SLy equation of state with equal mass $1.35\,\mathrm{M}_\odot$ progenitor neutron stars.
    The three panels have $\tc=5$\,ms (upper panels), $\tc=10$\,ms (centre panels), and $\tc=15$\,ms (lower panels).
    The left panels show the numerical-relativity injection (plus polarisation) in black and the posterior waveforms in blue.
    The vertical black lines show the beginning, $t=\tc$ (solid), and the end, $t=\tc+\tw$ (dashed), of the collapse for the injected signal.
    We perform injections into the 2A+ detector network at a grid of distances  and from these injections we select three distances where we can clearly recover the collapse time (see Sec.~\ref{sec:methodology}).
    The injection distances are 5.93\,Mpc for 5\,ms, 3.04\,Mpc for 10\,ms, and 1.00\,Mpc for 15\,ms.
    The right panels show posteriors, $p(\tc)$, in shaded blue, along with the true injected value, $\tc$, as solid vertical black lines.
    The model successfully recovers both the collapse time and the complex nature of the numerical-relativity injection, for all three injections.
    For reference, the full posteriors for $\tc=10\,$ms are shown in the appendix (Fig.~\ref{fig:FullCorner}).\par

\par
    In Fig.~\ref{fig:CollapseTimeDistanceGrid}, we show the maximum distance for which we can recover the collapse time for the post-merger remnant.
    The numerical-relativity injections are performed for SLy (upward triangles) and LS220 (downward triangles) equations of state.
    We inject into the following interferometers described in Sec.~\ref{sec:methodology}: 1) 2A+ (blue), 2) 2A+ and the proposed NEMO (orange), 3) Einstein Telescope (green), and 4) Einstein Telescope with Cosmic Explorer (red).
    We calculate the signal-to-noise ratio at a fixed distance over the entire sky for each detector network.
    We then choose a sky position with a signal-to-noise ratio close to the mean all-sky signal-to-noise ratio, and perform all injections at this sky position for this detector network.
    The lower error bars show the largest distance where the collapse times are recovered, the upper error bars show the smallest distance where the collapse time recoveries fail, and the marker is placed in the midpoint between these distances.
    If a post-merger collapse event occurs at a sky location near the antenna pattern maximum then the maximum distance where we can measure $\tc$ will increase by a factor of $\sim\!0.6$.
    \par
    
    Detecting the collapse time of a GW170817-like event at a luminosity distance of $40^{+8}_{-14}$\,Mpc~\cite{GW170817multi} (Fig.~\ref{fig:CollapseTimeDistanceGrid}, shaded region, gravitational wave only) would require the combination of ET with CE for $\tc \sim\!10$\,ms,  or ET with $\tc \sim\!5$\,ms. 
    The detection distance for 2A+ with  $\tc\sim\!10$\,ms is $\sim\!6\Hyphdash*8$\,Mpc.  
    This reduces to $\sim\!1.5\Hyphdash*2.0$\,Mpc for $\tc \sim\!15$\,ms. 
    Adding the NEMO high frequency detector increases the detection distance to $\sim\!17\Hyphdash*31$\,Mpc  for $\tc \sim\!10$\,ms.
    The detection distances for $\tc\sim\!15$\,ms with 2A+ and NEMO interferometers are $\sim\!3$\,Mpc and $\sim\!8$\,Mpc, for SLy and LS220 injections, respectively.\par

    For most collapse times and interferometer configurations, the ratio of the detection distance for LS220 to SLy injections is around $\sim\!1.2\approx f_0(\mathrm{SLy}) / f_0(\mathrm{LS220})$ ($f_0$ is the dominant post-merger oscillation frequency) which is consistent with SLy being softer and more compact than the LS220 equation of state. 
    For $\tc\!\sim15$\,ms injections into either 2A+ with NEMO, ET, or ET with CE, interferometer networks, the ratio of the LS220 to SLy detection distance increases to $\sim\!3$.
    Specifically, for $\tc\!\sim15$\,ms injections, detection distances of $\sim\!8,16,23$\,Mpc are found for LS220 equation of state with interferometers 2A+ with NEMO, ET, and, ET with CE, respectively.
    The corresponding detection distances for SLy injections are $\sim\!3,5,7$\,Mpc. 
    \par
    For injections where we can recover the collapse time, the dominant post-merger frequency is well constrained at $f_0 \lesssim 2\%$, $f_0 \lesssim 1\%$, and $f_0 \lesssim 0.2\%$ for injections of $\tc=5$\,ms, $\tc=10$\,ms and $\tc=15$\,ms, respectively. 
    Finally, we find no significant correlations between $\tc$ and other model parameters.
    We also attempt to measure the maximum detection distance where we can recover $\tc=20$\,ms and find that limitations in the third-order exponentially-damped sinusoidal model, Eq.~\ref{eq:hplus}, prevent recovery of such collapse times.
    For $\tc > 15$\,ms signals, the analytical model cannot successfully track the time-domain phase of the gravitational-wave strain.
    This limitation could be overcome by increasing the complexity of the frequency evolution in Eq.~\ref{eq:hplus}, possibly introducing a quadratic frequency evolution term~\cite{Bose2018}.
    Additionally, unmodelled searches such as \textsc{BayesWave} could be modified to measure the collapse time of the remnant~\cite{Cornish2015,Littenberg2015,Chatziioannou2017,Torres-Rivas2019}.\par
\vspace{-0.25cm}
\section{Discussion}\label{sec:discussion}

    We inject post-merger gravitational-wave signals that have been modified to collapse at varying distances into four different interferometer configurations: 2A+, 2A+ with NEMO, ET and ET with CE.
    We perform injections with collapse times of 5, 10, and 15\,ms, and recover collapse-time posteriors.
    The injected gravitational-wave strain is recovered with a third-order exponentially damped sinusoid with a linear frequency-drift term~\cite{Easter2020} that has been modified to collapse at $t=\tc$. \par
    
    To measure the collapse time of a post-merger remnant in a GW170817-like event (gravitational-wave only, luminosity distance of $40^{+8}_{-14}$\,Mpc~\cite{GW170817multi}), we find that we need interferometer configurations of either ET, or ET with CE, for $\tc \sim\!10$\,ms with the exclusion of ET with SLy equation of state.

    We show that, for each detector network, the maximum detection distance where we can measure 5\,ms collapse times is similar to the maximum detection distance for 10\,ms collapse times, with maximum detection distances of: $\sim\!6\Hyphdash*8$\,Mpc for 2A+, $\sim\!18\Hyphdash*26$\,Mpc for 2A+ with NEMO, $\sim\!23\Hyphdash*37$\,Mpc for ET, and $\sim\!33\Hyphdash*53$\,Mpc for ET with CE.\par
    
    We find that the stiffer equation of state, LS220, has more energy in the post-merger gravitational wave at larger times after the merger.
    This leads to larger maximum detection distances for LS220 equations of state relative to SLy injections for $\tc\sim\!15$\,ms.
    The maximum detection distance for each detector network for $\tc\sim\!15$\,ms are $\sim 1.5, 2.7, 5.4, 7.4$\,Mpc for SLy injections, and $\sim 2.0, 8.0, 16, 23$\,Mpc for LS220 injections, for 2A+, 2A+ with NEMO, ET, and ET with CE, detectors respectively.
    The above distances assume an injection at a sky position corresponding to an average signal-to-noise ratio over the entire sky.
    The detection distance would increase by a factor of $\sim\!0.6$ near an optimal sky position. \par
    
    We find that there are three predominant regions for detecting the collapse time. The first region, with small collapse times, is mainly limited by the Bayes Factor for the ratio of post-merger signal to noise. 
    For large collapse times, waveform systematics limit detections, specifically the inability of the model to track the phase of the gravitational-wave strain.
    Between these two regions the signal-to-noise ratio is the limiting factor.\par
    
    Ignoring waveform systematics, Ref.~\cite{Zhang2021} found that they could achieve a signal-to-noise ratio of 0.5\Hyphdash*8.6 for a collapse time of 10\,ms for a post-merger gravitational-wave signal at 50\,Mpc.
    The model used was a single-order damped sinusoid injected into a high-frequency detector.
    The authors used a TM1 equation of state with two equal mass 1.35\,M$_\odot$ progenitors with a dominant post-merger frequency of $\approx 2.8$\,kHz which very similar to $f_0$ for the LS220 equation of state in this paper.
    We find in this paper that we require a post-merger signal-to-noise ratio of $\gtrsim 17$ to successfully recover $\tc\sim\!10$\,ms for LS220 equation of state when waveform systematics are considered.
    
    With an estimated binary neutron star merger rate of $320^{+490}_{-240}\,\mathrm{Gpc^{-3}yr^{-1}}$~\cite{PopGWTC2}, it is unlikely that the collapse time of a post-merger remnant will be detected before either Cosmic Explorer or Einstein Telescope are operating at design sensitivity.
    When Cosmic Explorer and Einstein Telescope are both operating we may detect post-merger collapse times of $\sim\!10$\,ms.
    If only Einstein Telescope is fully operating then we may potentially measure post-merger collapse times of $\sim\!10$\,ms except for soft equations of state like SLy.
    In the mean time we will need to rely on indirect estimates of the post-merger collapse time that depend on multi-messenger observations~\cite[e.g.,][]{Metzger2018,Gill2019,Murguia-Berthier2020,Yu2018}.
    However, if a GW170817-like event occurred near an optimal sky position there would be a 60\% increase in the detection distance.
    In this case $\tc \sim\!10$\,ms may be detectable for ET, and ET with CE, for both equations of state.
    Additionally, 2A+ with NEMO would also be detectable for $\tc \sim 10$\,ms in this situation.
    It may also be possible to detune the proposed NEMO high frequency detector to increase sensitivity in the post-merger frequency band.
    This could potentially increase the sensitivity of the NEMO detector by a factor of $\sim\!1.6$ which would be enough to allow the NEMO detector with 2A+ to detect a GW170817-like post-merger collapse for $\tc \sim\! 10$\,ms. \par
    Finally, these results are dependent on the decay characteristics of the numerical-relativity simulations used in this paper.
    If the amplitude of future post-merger gravitational-waves have significantly longer decay timescales than the numerical-relativity simulations used here, then it is conceivable that larger collapse times could be measured.
    However, in this case waveform systematics become more important and models will either need to successfully track the waveform phase, or rely on incoherent methods that are independent of the phase of the gravitational-wave strain, or use unmodelled coherent detection methods~\cite[e.g.,][]{GW170817postmerger1,GW170817Postmerger2}.

\section{Acknowledgments}
        P.~D.~L. is supported through Australian Research Council (ARC) Future Fellowship FT160100112,  ARC Discovery Project DP180103155 and  ARC Centre of Excellence CE170100004. A.~R.~C. is supported in part by the Australian Research Council through a Discovery Early Career Researcher Award (DE190100656). Parts of this research were supported by the Australian Research Council Centre of Excellence for All Sky Astrophysics in 3 Dimensions (ASTRO 3D), through project number CE170100013. Parts of this work were performed on the OzSTAR national facility at Swinburne University of Technology. The OzSTAR program receives funding in part from the Astronomy National Collaborative Research Infrastructure Strategy (NCRIS) allocation provided by the Australian Government. The authors wish to acknowledge the public release of the Computational Relativity data at \url{http://www.computational-relativity.org}. We are grateful to Tim Dietrich for valuable comments on this manuscript.
\bibliography{CollapseTimeBib}

\appendix
\clearpage

\renewcommand\thefigure{\thesection.\arabic{figure}} 
\renewcommand\thetable{\thesection.\arabic{table}} 
\setcounter{figure}{0} 
\setcounter{table}{0}

\section{Priors}
\label{appendix:a}
    The priors are listed in Eqs.~\ref{eq:firstprior}\Hyphdash*\ref{eq:w01} with $\mathcal{U}(a,b)$ representing a uniform prior distribution from $a$ to $b$. 
    The mode number $j$ is limited to $\lbrace 0,1,2 \rbrace$. 
    The priors in Eqs.~\ref{eq:freqsorting}\Hyphdash*\ref{eq:w01} are constraining priors. 
    These restrictions are enforced in addition to the standard priors. 
    The priors in Eqs.~\ref{eq:freqsorting}\Hyphdash*\ref{eq:modesorting} sort the maximum spectral amplitude of each mode which improves computational stability and mode identification.
    See Ref.~\cite{Easter2020} for more details on mode sorting.
    We find that correlations between $\tc$ , $\tw$, and $T_0$ (the exponential decay time-constant for mode zero) make it very difficult to recover all three parameters simultaneously, even with analytic injections into zero noise.
    Fixing $\tw=2$\,ms allows recovery of all other parameters in both analytical injections with zero noise, and numerical-relativity injections with Gaussian noise.
    \vfill\null 
    \pagebreak
    
    \begin{eqnarray}
        \log_{10}{H} & \sim & \mathcal{U}(-24, -17)\label{eq:firstprior}  \\
        f_j & \sim & \mathcal{U}(1000,5000)\  [\mathrm{Hz}]\\
        \tc & \sim & \mathcal{U}(0,100)\  [\mathrm{ms}]\\
        \tw & = & 2\  [\mathrm{ms}]\\
        T_j & \sim & \mathcal{U}(0,100)\ [\mathrm{ms}]\\
        \psi_j & \sim & \mathcal{U}(-\pi,\pi) \\
        \alpha_j & \sim & \mathcal{U}(-6.4,6.4)\   [\mathrm{Hz}] \\
        w_0, w_1 & \sim & \mathcal{U}(0,1)\label{eq:lastprior}\\
        f_{j}  & > & f_{j+1}\label{eq:freqsorting}\\
        \max|\tilde{h}_{j}(f)|_f & > & \max|\tilde{h}_{j+1}(f)|_f \label{eq:modesorting}\\
        w_0+w_1 & < & 1  \label{eq:w01}\\
        w_2 & = & 1 - w_0 - w_1 \label{eq:w2}
    \end{eqnarray}
    \onecolumngrid
    \vfill\null 
    \pagebreak
    
\section{Example posteriors}
\label{appendix:b}
\begin{figure}[h]
        \centering
        \includegraphics[scale=0.5]{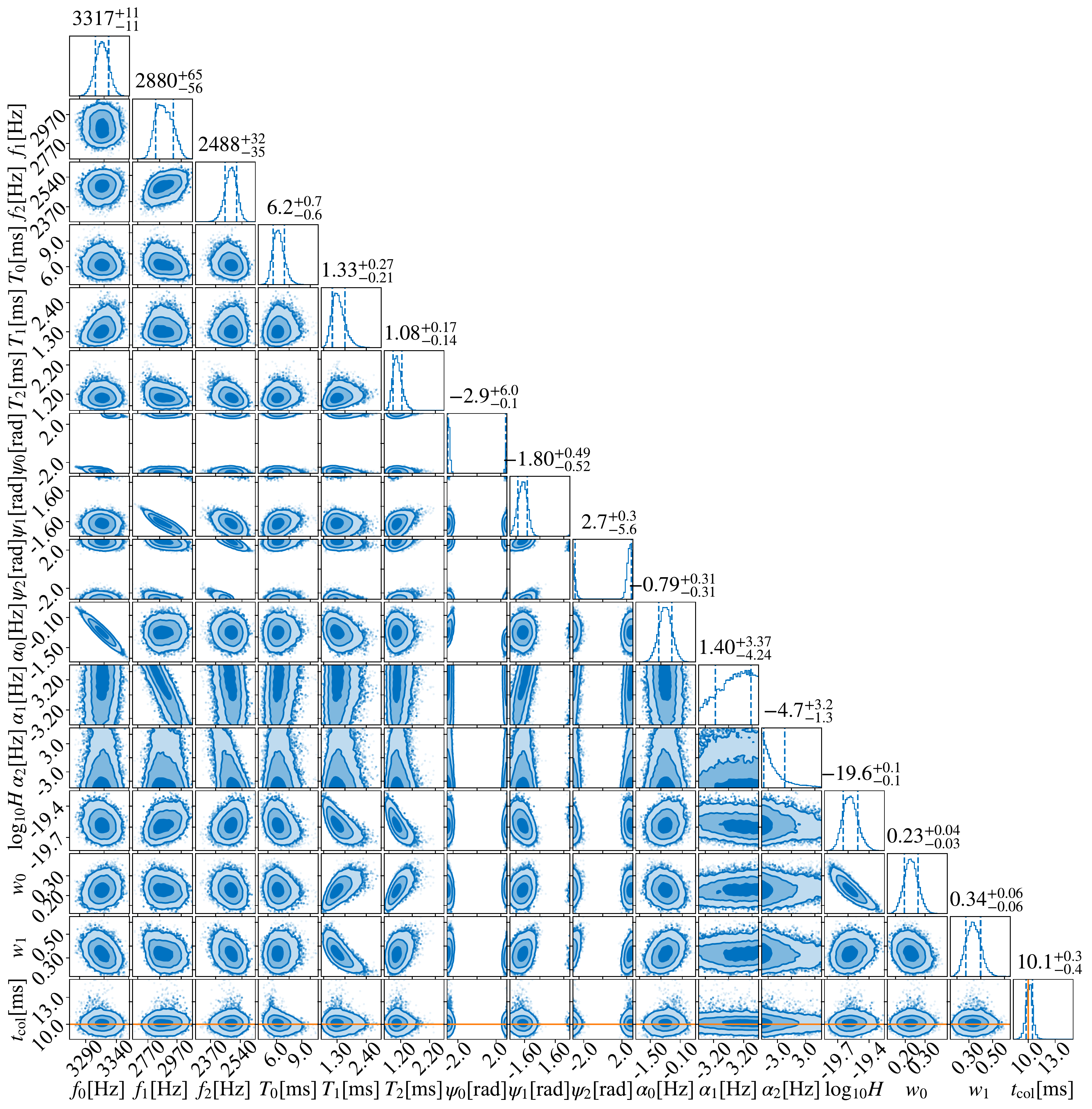}
        \caption
        {Full posteriors corresponding to the $\tc=10\,$ms injection shown in Fig.~\ref{fig:TimeResponse}. The  numerical-relativity simulation is injected into a 2A+ detector network at a distance of 3.04\,Mpc with an SLy equation of state and equal-mass $1.35\,\msun$ neutron stars. Credible intervals of 68\% are shown as dashed blue lines in the one-dimensional posteriors. The injected collapse time is shown by the orange lines in the lower-most panels.} 
        \label{fig:FullCorner}
\end{figure}

    \vspace{-0.3cm}
    Figure~\ref{fig:FullCorner} shows the posteriors for a post-merger numerical-relativity injection with $\tc=10\,$ms and SLy equation of state with equal mass $1.35\,\msun$ neutron stars. The injections are performed at a distance of 3.04\,Mpc into a detector network of 2A+.
    These posteriors correspond to the $\tc=10\,$ms time-domain signal and $p(\tc)$ in Fig.~\ref{fig:TimeResponse}.
    Orange lines on the bottom panels show the injected $\tc$.
    The recovered collapse time is $\tc=10.1^{+0.3}_{-0.4}\,$ms. 
    The primary post-merger oscillation frequency is $f_0=3317\pm 11\,$Hz with an exponential decay time-constant of $T_0=6.2_{-0.6}^{+ 0.7}\,$ms. 
    The linear frequency-drift term for the fundamental frequency is $\alpha_0=-0.79\pm 0.31\,$Hz. The frequencies corresponding to the sub-dominant modes are $f_1=2880^{+65}_{-56}\,$Hz and $f_2=2488^{+32}_{-35}\,$Hz. 

\vspace{-0.5cm}

\end{document}